\documentclass[11pt]{article}

\author{Pedro F. da Silva J{\'u}nior\footnote{Correspondence email: pedro.fsilva2@ufpe.br}}

\usepackage[english]{babel}
\usepackage[utf8]{inputenc} 
\usepackage[a4paper,top=2.0cm,bottom=2.0cm,left=2.0cm,right=1.2cm,columnsep=0.6cm]{geometry}
\usepackage{amsmath} 
\usepackage[makeroom]{cancel}
\usepackage{titlesec}
\titlelabel{\thetitle.\quad} 
\usepackage{multicol} 
\usepackage{enumitem} 
\usepackage{epsfig,graphics,graphicx} 
\usepackage{subfigure}
\usepackage[font=small,labelfont=bf]{caption}
\usepackage{xcolor} 
\PassOptionsToPackage{hyphens}{url}
\usepackage[colorlinks=true,urlcolor=blue,allbordercolors={0 0 0},pdfborderstyle={/W 1}]{hyperref}
\usepackage{amssymb}
\DeclareMathAlphabet{\altmathcal}{OMS}{cmsy}{m}{n}
\usepackage{mathptmx}
\usepackage{pifont}
\usepackage{bibentry}

\makeatletter

\newenvironment{figurehere}
  {\def\@captype{figure}}
  {}
\makeatother

\newcommand*{\QEDA}{\null\nobreak\hfill\ensuremath{\blacksquare}}

\let\oldquote\quote
\let\endoldquote\endquote
\renewenvironment{quote}[2][]
  {\if\relax\detokenize{#1}\relax
     \def\quoteauthor{#2}%
   \else
     \def\quoteauthor{#2~---~#1}%
   \fi
   \oldquote}
  {\par\nobreak\smallskip\hfill(\quoteauthor)%
   \endoldquote\addvspace{\bigskipamount}}

\title{\bf {On the Deduction of the Carathéodory's Axiom of the Second Law of Thermodynamics from the Clausius and Kelvin Principles}}
\date{\vspace{-5ex}}

\begin{document}

\maketitle 

\vspace{-0.5cm}

\begin{center}

\small{\textit{Universidade Federal de Pernambuco, Caruaru, PE, Brazil}}

\vspace{0.5cm}

\small{\textbf{Abstract}}

\vspace{0.5cm}

\begin{minipage}[l]{14cm} {\small

Carathéodory’s formalism for classical thermodynamics is a rich alternative approach to this theory, although unpopular with students and physics professors. This approach dispenses with the content of thermal machines for the presentation of the second law of thermodynamics. In this paper, we discuss Carathéodory's formalism historically, and show how Carathéodory's axiom of the second law of thermodynamics is derived, didactically, from the Clausius principle and the Kelvin principle. In addition, also providing an objective character for this paper, in the sense of seeking to popularize the teaching of Carathéodory's formalism in disciplines of classical thermodynamics at undergraduate level, we guide the reader to obtain the entropy and mathematical content of the second law of thermodynamics through this formalism. Finally, considering the wide reviewed literature, the proof we gave for deducing the Carathéodory's axiom of the second law of thermodynamics from the Clausius principle is new.}

\end{minipage}

\end{center}

\vspace{0.3cm}

\begin{multicols}{2}

\section{Introduction}\label{sec:introducao}

In physics, we are often presented with distinct but equivalent conceptual and mathematical approaches to the same theory. By equivalent we mean that these different descriptions for the same theory obtain, without any loss of physical content, the same final results. Moreover, generally, the paths and methods used by each description differ enormously from each other. These distinct descriptions for the same theory we call \textit{formalisms}. A famous case of formalisms in physics occurs in classical mechanics, where we have the formalisms due to Newton, Lagrange, and Hamilton.

Another discipline of physics that allows the use of formalisms is classical thermodynamics. Classical thermodynamics is the perspective of thermodynamics that studies physical systems from laws that generalize the observations made about the \textit{macroscopic} behavior of these systems. To do this, classical thermodynamics ignores the \textit{microscopic} nature of matter. This differs from another famous perspective of thermodynamics that considers for its description the \textit{microscopic} nature of matter and advances in \textit{statistical mechanics}. This other perspective on thermodynamics is \textit{statistical thermodynamics}. It is not statistical thermodynamics that we will be dealing with here.

Classical thermodynamics is a highly solidified and well-established area of physics throughout the scientific community. Moreover, it is a theory whose content is considerably popular, from the most basic levels of science education, to higher level courses in physics and related areas, such as chemistry, engineering, etc. Thus, the fundamental theoretical material of classical thermodynamics should cause little strangeness to the reader. Specifically about the existing formalisms for classical thermodynamics, and also about its own teaching at the undergraduate level, it is notable the traditional presentation of this subject from the perspective of the efficiency of thermal engines -- or thermal machines. 

Majority in current textbooks, this traditional formalism makes use of the two equivalent experimental principles of the second law of thermodynamics -- the Clausius and Kelvin principles -- in conjunction with Carnot's theorem of thermal machines, to obtain the \textit{entropy} and the mathematical content of the \textit{second law of thermodynamics} -- the principle of entropy increase. This traditional formalism was built by Clausius on historical and technical developments by names like Carnot, Clapeyron, and Kelvin \cite{muller2007}. Thus, we name this traditional formalism here as \textit{Clausius's formalism}, inspired also by the teaching literature of classical thermodynamics \cite{oliveira2017}. 

However, as we have previously announced, the Clausius's formalism is not the only possible one to describe classical thermodynamics. Another famous formalism originated in the work of Gibbs \cite{gibbs1928} who, in a series of papers between 1873 and 1878, advocated analytical methods for describing classical thermodynamics. In doing so, Gibbs influenced several popular works of classical thermodynamics that appeared later \cite{callen1985,guggenheim1986}. These works constructed a classical thermodynamics of postulates, introducing fundamental thermodynamic notions such as \textit{entropy} in the form of elementary concepts from which the other concepts of the theory are derived \cite{oliveira2019}. In general, we call here \textit{Gibbs's formalism} the formalisms due to the pioneering work of Gibbs.    

Besides these, there is also the \textit{Carathéodory's formalism} for classical thermodynamics, which is the formalism that most interests us in this paper. The Carathéodory's formalism differs from the other two formalisms mentioned above in that it does not need the thermal machines used in the Clausius's formalism and, although it also makes use of postulates, or axioms, it does not do so using the same methodology involved in the Gibbs's formalism. The Carathéodory's formalism plays an important role in the construction of the theory of classical thermodynamics itself, as the following quote tells:

\begin{quote}[p. 149]{\cite{luscombe2018}}
Entropy was discovered by a somewhat circuitous path through the efficiency of heat engines, a finding that in hindsight could appear serendipitous. Were we just {\it{lucky}} to have discovered something so fundamental in this way? Can it be seen {\it{directly}} that entropy as a state variable is contained in the structure of thermodynamics, without the baggage of heat engines? It can, as shown
by Constantin Carathéodory in 1909.
\end{quote}

These are the first words of James H. Luscombe in the introduction to the tenth chapter of his recent\footnote{Luscombe presents in his book a modern approach to the Carathéodory formalism, working it with the concept of vector fields.} \textit{Thermodynamics} \cite{luscombe2018}. Despite its recognized importance\footnote{Carathéodory's work is a milestone in terms of the descriptive foundations of an analytic classical thermodynamics \cite{thess2011,lieb1999}.} Carathéodory's formalism is almost completely unknown today by most physics professors and students. Examples of didatic productions that teach the Carathéodory's formalism are also currently scarce. That said, we present in the \ref{sec:formalismo} section a short historical discussion of the background and methods of Carathéodory's formalism. We also show, in sections \ref{sec:kelvin}, \ref{sec:clausius}, and \ref{sec:entropia}, how Carathéodory's formalism connects with one of the most important results of classical thermodynamics: the second law of thermodynamics, formulated from the principles of Clausius and Kelvin.

\section{Carathéodory's formalism} \label{sec:formalismo}

In a 1909 paper published in the \textit{Mathematische Annalen}, the mathematician Constantin Carathéodory \cite{caratheodory1909}, proposed a formalism for classical thermodynamics that obtained the results of the theory\footnote{The full expected results of thermodynamics do not arise from Carathéodory's formalism. The same is also true of the Clausius's formalism. The greatest example of this is the third law of thermodynamics, whose precise and complete content can only be obtained by the advent of quantum mechanics. Without this inclusion, both formalisms produce the content of the third law of thermodynamics in a {textit{ad hoc}} fashion. In this sense, we can say that these formalisms provide what would be expected as the classical mathematical content of the theory. For more details on this subject, we suggest that the reader consult Tisza's book \cite{tisza1966}.} starting with two axioms, one for the first law and the other for the second law of thermodynamics, in such a way that the development of thermodynamic concepts took place in terms of considerations arising from mechanical concepts.

The axiom used by Carathéodory for the second law of thermodynamics was the greatest innovation in his work.  This axiom was not based on experiments, nevertheless, from it emerged the \textit{entropy} and its \textit{classical mathematical content}. Although this axiom is eventually known as the \textit{second Carathéodory's axiom} \cite{sears1966}, given the existence of a Carathéodory's axiom also for the first law of thermodynamics, our major focus here is the study in particular of the \textit{second law of thermodynamics}. For this reason, we will henceforth refer to Carathéodory's axiom for the second law of thermodynamics only as \textit{Carathéodory's axiom}.

Several authors have dedicated themselves to the mission of disseminating Carathéodory's ideas over the years, namely: Sears \cite{sears1966}, Chandrasekhar \cite{chandrasekhar1939}, Buchdahl \cite{buchdahl1966}, Landsberg \cite{landsberg2014}, Dunning-Davies \cite{davies2007}, among others. In particular, Max Born, one of the forerunners of quantum mechanics, defended Carathéodory's view for a large part of his life. With publications \cite{born1921,born1949} and also with harsh criticism of Clausius's formalism, Born sought to popularize Carathéodory's formalism by saying, for example, in a 1921 paper:

\begin{quote}
{\cite{born1921}}
{\textit{a) There is no other area of physics where considerations are applied that bear any resemblance to
the Carnot cycle and its correlatives. b) One has to admit that
thermodynamics, in its traditional model, has not yet realized the logical ideal of separation between the physical content
and the mathematical description. c) It is necessary to do a removal
of rubble, which a tradition full of too much piousness
hitherto, has not dared to remove.}}
\end{quote}

In correspondence to Einstein \cite{correspondencia}, exactly about this publication \cite{born1921}, that he submitted to the current \textit{Physikalische Zeitschrift}, Born wrote: 

\begin{quote}[p. 53]{\cite{correspondencia}}
\begin{flushright}
Frankfurt a.M.\\
12 February, 1921
\end{flushright}
Dear Einstein\\
... I have done little theoretical work. I have recently written an account of Carathéodory's thermodynamics, which will appear
shortly in the Physikalische Zeitung. I am very curious to know
what you will say about it.
\end{quote} 

Having already published the papers that would elevate his name to the rank of one of the greatest in the history of science - on relativity and the photoelectric effect - positive feedback from Einstein on the Carathéodory's formalism could certainly have given this matter a different reception in the physics community. But, despite Born's request, there is no mention in the subsequent correspondences between Born and Einstein \cite{correspondencia} of the latter's concern over Carathéodory's work. Afterwards, and still on his attempt to popularize the work of Carathéodory, the frustration of Born is confirmed when he states, referring by \textit{classical method} to what we here call Clausius's formalism:

\begin{quote}[p. 55]{\cite{correspondencia}}
My interpretation of Carathéodory's thermodynamics did not have the
effect I had hoped for of displacing the classical method which, in my opinion, is both clumsy and mathematically opaque.
\end{quote}

After Born, possible obstacles to the use and dissemination of Carathéodory's formalism were investigated by various other authors. In this context, besides the apparent historical bad luck that we have cited, pedagogical and mathematical obstacles that may have contributed to the unpopularity of the Carathéodory's formalism were pointed out by Zemansky \cite{zemansky1966} in 1966. Zemansky wrote that for the presentation of the second law of thermodynamics: 

\begin{quote}[p. 915]{\cite{zemansky1966}}
Due to the fact that Carathéodory's axiom was not based directly on experience and that the proof of his theorem was longwinded and difficult, most physicists and textbook writers ignored the Carathéodory treatment [...]
\end{quote} 

Curiously, prior to Zemansky's account there were already papers that solved the issues he pointed out. For example, in 1964, Landsberg \cite{landsberg1964} proved that the \textit{Carathéodory's axiom} can be deduced from Kelvin's principle. This result was further investigated by Titulaer and Van Kampen \cite{titulaer1965} a year later, in 1965. In that same year, Dunning-Davies \cite{davies1965} proved the reciprocal of Landsberg's conclusion, establishing the \textit{equivalence between Carathéodory's axiom and Kelvin's principle}. So, even though it is not based directly from experimental facts, Carathéodory's axiom was shown to be equivalent to Kelvin's experimental principle, so that Carathéodory's axiom can be seen as just another statement of the second law of thermodynamics \cite{buchdahl1966}.

On the other hand, the \textit{Carathéodory's theorem} was made demonstrable with short arguments, related to the geometry of the \textit{thermodynamic space} \cite{buchdahl1966,born1921,born1949}, as well as, related to the use of the concept of vector fields \cite{luscombe2018}. These demonstrations, however, follow a level of simplicity that does not contemplate all the mathematical content of Carathéodory's theorem in its general version, as Boyling \cite{boyling1968} showed. Nevertheless, for the teaching of classical thermodynamics, the justifications given by the aforementioned authors in the demonstration of this theorem are, as they propose to be, sufficient, and do not impair the following of the physical results of the theory \cite{buchdahl1966}.

As is already clear, overcoming these two difficulties pointed out by Zemansky to the Carathéodory's formalism was not enough to make this formalism spread later in physics.  But then, were there any other major difficulties, besides those originally indicated by Zemansky, with the use of the Carathéodory's formalism? And in particular in the context of the presentation of the second law of thermodynamics? We argue with this paper that no. And so, we aim to introduce some of the methods and meanings of the Carathéodory's formalism to the reader, specifically in the context of the presentation of the second law of thermodynamics.

Therefore, in the following sections, we seek to didactically introduce the reader to the efforts we have cited related to connecting the Carathéodory's axiom with the Clausius principle and the Kelvin principle. As a novel result from the extensive literature reviewed in this paper, we prove the \textit{direct deduction of the Carathéodory's axiom from the Clausius principle}. The section \ref{sec:kelvin} deals with the deduction of the Carathéodory's axiom from Kelvin principle, according to Titulaer and Van Kampen \cite{titulaer1965}. The section \ref{sec:clausius}, on the other hand, deals with our deduction of the Carathéodory's axiom from the Clausius principle. We consider these steps to be the most important to the reader with regard to the connection of Carathéodory's formalism with the second law of thermodynamics.

Next, seeking to contribute in some measure to the popularization of Carathéodory's formalism in classical thermodynamics courses at the undergraduate level, we show to the reader in the section \ref{sec:entropia} a glimpse of how the \textit{entropy} and the mathematical content of the \textit{second law of thermodynamics} arise as a direct consequence of the application of Carathéodory's theorem. To this end, Carathéodory's theorem was demonstrated in the subsection \ref{subsec:teorema} in a simple way, from Born's argument \cite{born1921}. The \textit{entropy} itself, on its turn, was covered soon after, in subsection \ref{subsec:segundalei}. 

Also, for a good understanding of what follows, a basic knowledge of the fundamental concepts of classical thermodynamics is sufficient for the reader: \textit{system and neighborhood; thermal reservoir, equilibrium and thermodynamic space; coordinates, state and thermodynamic processes; first and second law of thermodynamics, etc}. However, whenever necessary, for the sake of emphasis, we will briefly discuss some of these concepts. 

Finally, this paper is not intended to defend the superiority of the Carathéodory's formalism with respect to the other formalisms of classical thermodynamics. Hence, we only try here to indicate the possibility of using the Carathéodory's formalism in the teaching of classical thermodynamics.

\section{Carathéodory's axiom from Kelvin's principle} \label{sec:kelvin} 

There are some differences in the literature regarding the writing of the experimental principles of the second law of thermodynamics. Thus, even if merely related to a slightly different choice of words by each author, these differences can cause confusion in the interpretation of the statements of these principles \cite{davies2011}. In an effort to avoid such situations, this paper will state both the Carathéodory's axiom and the Clausius and Kelvin principles as found in the classic book \textit{An Introduction to the Study of Stellar Structure} by the 1983 Nobel Prize in Physics winner Subrahmanyan Chandrasekhar.

It then follows Kelvin's principle, according to Chandrasekhar \cite{chandrasekhar1939}: 

\begin{quote}[p. 24]{\cite{chandrasekhar1939}}
\textit{In a cycle of processes it is impossible to transfer heat from a heat reservoir and convert it all into work, without at the same time transferring a certain amount of heat from a hotter to a colder body.} 
\end{quote}

What Kelvin's principle -- which will be referred to hereafter for short as (K) -- says is that, during any thermodynamic cycle, it is not possible for a system to fully convert heat $\altmathcal{Q}$ absorbed from a thermal reservoir into work $W$ without, during the same cycle, there also being heat given off from the system to another system at a lower temperature. In other words, let $\altmathcal{Q}$ be the heat absorbed by a system from a thermal reservoir during any thermodynamic cycle, and let $W$ be the work related to the interaction of the system with its neighborhood during that cycle. Then, by (K), the following equality at the end of the cycle is impossible, with $\altmathcal{Q}>0$

\begin{equation}
\altmathcal{Q} = W.
\label{1}
\end{equation}

Observe that (K) prominently works with the concepts of heat, work, and temperature, in an \textit{a priori} fashion. That is, in this statement of the second law of thermodynamics, heat, work, and temperature are elementary thermodynamic concepts.

Understood (K), the Carathéodory's axiom follows, also according to Chandrasekhar \cite{chandrasekhar1939}: 

\begin{quote}[p. 24]{\cite{chandrasekhar1939}}
\textit{Arbitrarily near to any given state there exist states which cannot be reached from an initial state by means of adiabatic processes.} 
\end{quote}

There is much to comment on the Carathéodory's axiom -- which will be referred to hereafter, for short, as (AC) -- but initially it is necessary to understand the word \textit{state} in this context. Classical thermodynamics deals with macroscopic systems in \textit{equilibrium situations}, that is, in situations where the coordinates, or variables, of the system are well defined. Indeed, when \textit{equilibrium} is established, analogously to the case of classical mechanics, we have the complete characterization of the thermodynamic system by the value of its independent thermodynamic coordinates. And, when this is done, the state of the system is defined and is expressed by the set of these independent thermodynamic coordinates that characterize the equilibrium. So, for classical thermodynamics, \textit{situations of equilibrium are equivalent to defining the state of the thermodynamic system}.  

That said, let's get to the content of (AC). What (AC) states is that, given any particular state of a thermodynamic system, there will be other states that the system cannot reach through adiabatic processes. Adiabatic processes are processes that occur without energy exchange in the form of heat between the system and its surroundings. Note that (AC) makes no distinction between reversible or irreversible processes. Notice here that (AC) assumes as a major elementary thermodynamic concept the notion of an adiabatic process. This construction is present in Carathéodory's formalism in order to get away from the direct concept of heat flow, exchanging it for the idea of \textit{processes not purely mechanical}\footnote{For an extensive treatment of this question, we suggest the reader consult section 8 of Buchdahl's book \cite{buchdahl1966}.}.

But, to really understand the statement of (AC) we need to analyze the meaning of the term ``reach'' in it. To get from one state to another, the thermodynamic system needs to perform a process and then reach a new equilibrium situation, thus configuring a new thermodynamic state. To say that there are states that cannot be ``achieved'' by adiabatic processes means to say that there are equilibrium situations that cannot be achieved if we use an adiabatic process as a way to do so. Put another way, \textit{given any state of a thermodynamic system, we cannot subject the system to arbitrary adiabatic processes}.

We now seek to show that (K) $\Rightarrow$ (AC). To do this we need only show that if (AC) is false, then (K) is also\footnote{In other words, if it is true that the negation of (AC) provides the negation of (K), then since (K) is true, (AC) will also be so.}. Then, according to Titulaer and Van Kampen \cite{titulaer1965}, we first prove that (K) $\Rightarrow$ (AC) for \textit{reversible processes}. Reversible processes are, as the name suggests, processes that are amenable to being executed in both ``time directions'', since all situations of the system during these processes are equilibrium situations. For example, if we can conceive of a system consisting of a gas confined in a container bounded by a frictionless moving piston, we can lower the piston by depositing grains of sand one by one onto it, thereby reversibly compressing the gas in the container, so that the gas will always remain in equilibrium in this process. On the other hand, if we remove the grains of sand from the piston one by one, we may, at some point, return to the exact same equilibrium situation that we initiated this reasoning, returning the gas to its original thermodynamic state. This reversible temporal behavior defines reversible processes. Therefore, reversible processes are represented by continuous curves in thermodynamic space.

Thus, let be a thermodynamic space defining the thermodynamic states for a model thermodynamic system with a usual set of three independent thermodynamic coordinates: $\theta$, the \textit{empirical temperature} of the system, measured by some measuring instrument, on some temperature scale; $x_1$ and $x_2$, two coordinates related to the mechanical behavior of the thermodynamic system. For example, $x_1 = V$ and $x_2 = M$ being respectively the \textit{volume} and \textit{magnitude of magnetization} of the system. The choice of three independent coordinates for the characterization of the model thermodynamic system under discussion is simply because of the initial facility of working in three dimensions. However, naturally, what will be argued below also holds for a larger number of thermodynamic coordinates.

This thermodynamic space, represented in Fig.$\,$\ref{f1}, characterizes any thermodynamic state by the measured values of the coordinates $({\theta},{x_1}, {x_2})$, for example: $({{\theta}^0},{{x_1}^0}, {{x_2}^0})$. Consequently, such a state is uniquely identified as a point $A$ in said space such that, $A = ({{\theta}^0},{{x_1}^0}, {{x_2}^0})$.

\begin{figurehere}
\begin{center}	
\includegraphics[scale=0.68]{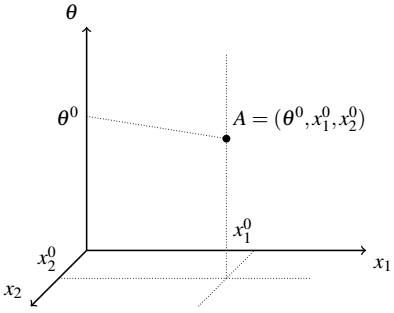}
\caption{Thermodynamic space of coordinates $\theta$, $x_1$ and $x_2$.}
\label{f1}
\end{center}
\end{figurehere}

Moreover, because we are dealing with \textit{thermodynamic systems}, it is always possible to choose the \textit{empirical temperature} as one of the independent thermodynamic coordinates. Next, let $\altmathcal{P}$ be the following \textit{reversible cycle} given in the thermodynamic space of Fig.$\,$\ref{f1} and represented in Fig.$\,$\ref{f2}.

\begin{figurehere}
\begin{center}	
\includegraphics[scale=0.7]{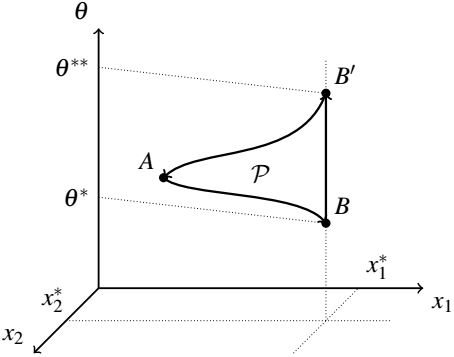}
\caption{Reversible cycle $\altmathcal{P}$ constructed by intermediate processes between points $A$, $B$ and $B'$ in thermodynamic space. Since $\altmathcal{P}$ is reversible, the direction of travel in the cycle can be chosen arbitrarily. Here the direction $A \rightarrow B \rightarrow B' \rightarrow A$ is chosen.}
\label{f2}
\end{center}
\end{figurehere}

The \textit{reversible cycle} $\altmathcal{P}$ of Fig. $\,$\ref{f2} is set up such that: the process $A \rightarrow B$ is assumed to be adiabatic, so ${\altmathcal{Q}}_{A \rightarrow B} = 0$; the process $B \rightarrow B'$ does not escape the line that preserves the values of ${x_1} = {x_1}^*$ and ${x_2} = {x_2}^*$, so there is no work involved from $B$ to $B'$, however there is heat absorbed ${\altmathcal{Q}}_{B \rightarrow B'} > 0$ from $B$ to $B'$ due to the temperature difference between $B$ and $B'$, ${\theta}^{**} - {\theta}^{*} > 0$; finally, the process $B' \rightarrow A$ is also supposed to be adiabatic, so ${\altmathcal{Q}}_{B' \rightarrow A} = 0$. This closes the ${\altmathcal{P}}$ cycle. For this cycle we have constructed it is important to realize that since $\altmathcal{P}$ is reversible, this whole idealized construction for $\altmathcal{P}$ could be reversed by inverting the cycle and taking $B' \rightarrow B$ in reverse such that, in this sense, there would be heat given up ${\altmathcal{Q}}_{B' \rightarrow B} < 0$ from $B'$ to $B$.

However, a more careful look at the $\altmathcal{P}$ cycle we have constructed reveals that if it is possible, then (AC) is false. Indeed, one can approximate $B'$ to $B$ abitrarily. Moreover, since $\altmathcal{P}$ is reversible, both the adiabatic process $A \rightarrow B$ and the adiabatic process $A \rightarrow B'$ can be realized. But if $B'$ is approximated arbitrarily from $B$, and from $A$ both $B$ and $B'$ can be reached by reversible adiabatic processes, then from $A$ all states on the same line ${{x_1}^*}{{x_2}^*}$ of $B$ and $B'$ can be reached by reversible adiabatic processes. Hence, also approximating the line ${{x_1}^*}{{x_2}^*}$ arbitrarily close to $A$, we would have that: \textit{arbitrarily close to $A$ there are states that can be reached from $A$ by reversible adiabatic processes, thus falsifying (AC) for reversible processes}.

But since we have so far no real -- physical -- justifications for the validity of (AC), let us suppose that $\altmathcal{P}$ is indeed possible and then (AC) is indeed false. Let us now apply the first law of thermodynamics to $\altmathcal{P}$. This gives us, thanks to the additivity of energy

\begin{equation}
\Delta{E_{\altmathcal{P}}} = \Delta{E_{A \rightarrow B}} + \Delta{E_{B \rightarrow B'}} + \Delta{E_{B' \rightarrow A}}.
\label{2}
\end{equation}

Since $\altmathcal{P}$ is a thermodynamic cycle, $\Delta{E_{\altmathcal{P}}} = 0$. Applying the characteristics of $\altmathcal{P}$ to (\ref{2}), we have

\begin{equation}
- W_{A \rightarrow B} + {\altmathcal{Q}}_{B \rightarrow B'} - W_{B' \rightarrow A} = 0.
\label{3}
\end{equation}

Note that the effective algebraic contributions of the quantities of work that appear in $\altmathcal{P}$ are related to the realization of work of the system in the neighborhood, or of the neighborhood on the system. Then, the expression (\ref{3}) can be rearranged. Naming $W$ the \textit{effective work} involved in the path of $\altmathcal{P}$, we get

\begin{equation}
\altmathcal{Q} = W.
\label{4}
\end{equation}

But this is analogous to what equation (\ref{1}) says, and so equation (\ref{4}), which comes from the assumption that (AC) is false, falsifies (K). Hence, for \textit{reversible} processes (K) $\Rightarrow$ (AC). However, we know that we also deal with \textit{irreversible} processes in classical thermodynamics. These processes are, as the name suggests, processes that can only be realized in a single ``time direction'', since the intermediate situations of the system in an irreversible process are not equilibrium situations. Even though classical thermodynamics is a theory that studies only equilibrium situations, in it a qualitative analysis of irreversible processes is also possible. This is due to the fact that in classical thermodynamics the initial and final situations of irreversible processes \textit{are always equilibrium situations}. Furthermore, it is from irreversible processes that the true meaning of \textit{entropy} and \textit{second law of thermodynamics} emerges.

In terms of thermodynamic space, irreversible processes cannot be represented as the usual continuous curves in this space, unlike reversible processes. Thus, since only the initial and final states of an irreversible process are defined, it is usual to represent it as a dashed line in thermodynamic space, connecting its initial and final states. It is natural that we try to evaluate the relation (K) $\Rightarrow$ (AC) for irreversible processes as well. So let be the \textit{irreversible cycles} ${\altmathcal{P}}_1$ and ${\altmathcal{P}}_2$ in the same previous thermodynamic space as Fig.$\,$\ref{f1}, represented, respectively, in Fig.$\,$\ref{f3} and Fig.$\,$\ref{f4}.

\begin{figurehere}
\begin{center}	
\includegraphics[scale=0.7]{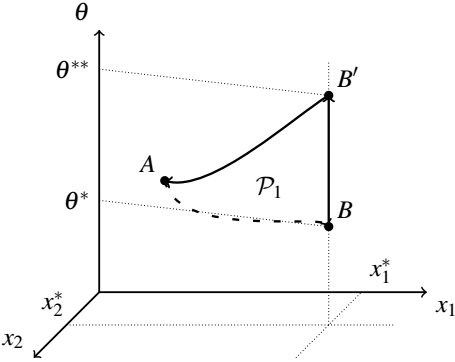}
\caption{Irreversible cycle ${\altmathcal{P}}_1$ constructed through intermediate processes between points $A$, $B$ and $B'$ in thermodynamic space. The intermediate process $A \rightarrow B$ is irreversible. Therefore, ${\altmathcal{P}}_1$ can only be realized in the direction $A \rightarrow B \rightarrow B' \rightarrow A$.}
\label{f3}
\end{center}
\end{figurehere}

\begin{figurehere}
\begin{center}	
\includegraphics[scale=0.7]{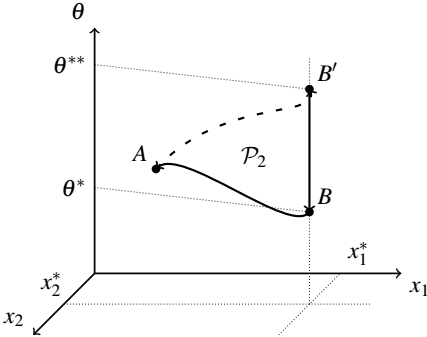}
\caption{Irreversible cycle ${\altmathcal{P}}_2$ constructed through intermediate processes between points $A$, $B$ and $B'$ in thermodynamic space. The intermediate process $A \rightarrow B'$ is irreversible. Therefore, ${\altmathcal{P}}_2$ can only be realized in the direction $A \rightarrow B' \rightarrow B \rightarrow A$.}
\label{f4}
\end{center}
\end{figurehere}

The ${\altmathcal{P}}_1$ and ${\altmathcal{P}}_2$ cycles are constructed analogously to that posed for the ${\altmathcal{P}}$ cycle in the reversible case. In the cycle ${\altmathcal{P}}_1$: the irreversible process $A \rightarrow B$ is assumed to be adiabatic, so ${\altmathcal{Q}}_{A \rightarrow B} = 0$; the reversible process $B \rightarrow B'$ holds on the line where ${x_1} = {x_1}^*$ and ${x_2} = {x_2}^*$, so $W_{B \rightarrow B'} = 0$, however ${altmathcal{Q}}_{B \rightarrow B'} > 0$ thanks to the temperature difference between $B$ and $B'$, ${\theta}^{**} - {\theta}^{*} > 0$; finally, the reversible process $B' \rightarrow A$ is also supposed to be adiabatic, so ${\altmathcal{Q}}_{B' \rightarrow A} = 0$. This closes the cycle ${\altmathcal{P}}_1$. It should be pointed out that thanks to the \textit{irreversibility} of ${\altmathcal{P}}_1$ it can only be traversed in the direction $A \rightarrow B \rightarrow B' \rightarrow A$.

Similarly, in the cycle ${\altmathcal{P}}_2$: the irreversible process $A \rightarrow B'$ is assumed to be adiabatic, so ${\altmathcal{Q}}_{A \rightarrow B'} = 0$; the reversible process $B' \rightarrow B$ holds on the line where ${x_1} = {x_1}^*$ and ${x_2} = {x_2}^*$, so $W_{B' \rightarrow B} = 0$, however ${\altmathcal{Q}}_{B' \rightarrow B} < 0$ thanks to the temperature difference between $B'$ and $B$, ${\theta}^{*} - {\theta}^{**} < 0$; lastly, the reversible process $B \rightarrow A$ is also supposed to be adiabatic, so ${\altmathcal{Q}}_{B \rightarrow A} = 0$. This closes the cycle ${\altmathcal{P}}_2$. Similarly as for the cycle ${\altmathcal{P}}_1$, thanks to the \textit{irreversibility} of ${\altmathcal{P}}_2$ it can only be traversed in the direction $A \rightarrow B' \rightarrow B \rightarrow A$.

Now suppose that both irreversible cycles ${\altmathcal{P}}_1$ and ${\altmathcal{P}}_2$ are simultaneously possible, i.e., both cycles can be performed. It turns out that if this is the case, then (AC) is false. In effect, again, we approximate $B'$ to $B$ in an arbitrary way. Moreover, if we suppose ${\altmathcal{P}}_1$ and ${\altmathcal{P}}_2$ to be simultaneously possible, then as a consequence both irreversible processes $A \rightarrow B$ and $A \rightarrow B'$ are also possible. Then, if $B'$ is approximated arbitrarily from $B$, and from $A$ one can reach both $B$ and $B'$ by irreversible adiabatic processes, then from $A$ all states on the same line ${{x_1}^*}{{x_2}^*}$ as $B$ and $B'$ can be reached by irreversible adiabatic processes. Hence, also approximating the line ${{x_1}^*}{{x_2}^*}$ arbitrarily close to $A$, we would have that: \textit{arbitrarily close to $A$ there are states that can be reached from $A$ by irreversible adiabatic processes, thus falsifying (AC) for irreversible processes}.

At this point, compared to the argument for the reversible case, the attentive reader should have already figured out what the next step is to be. Again, in principle we have no physical argument that prevents (AC) from being false in the irreversible case. So suppose that (AC) is really false and then ${\altmathcal{P}}_1$ and ${\altmathcal{P}}_2$ are simultaneously possible. Then, repeating the argument for the reversible case and applying the first law of thermodynamics to both ${\altmathcal{P}}_1$ and ${\altmathcal{P}}_2$, we find that, at the end of each of these cycles

\begin{equation}
\altmathcal{Q} = W.
\label{5}
\end{equation}

In the expression (\ref{5}) $\altmathcal{Q}$ is the heat absorbed, or ceded, by the system in interaction with an appropriate thermal reservoir for each of the cycles, and $W$ is the \textit{effective work} related to the interactions of the system with its neighborhood also for each of the cycles. Assuming that ${\altmathcal{P}}_1$ and ${\altmathcal{P}}_2$ are simultaneously possible, it is clear that an equality analogous to the expression (\ref{5}) could be written for each of these irreversible cycles: one in which $W = \altmathcal{Q} > 0$ at the end of the cycle, related to cycle ${\altmathcal{P}}_1$, and one in which $W = \altmathcal{Q} < 0$ at the end of the cycle, related to cycle ${\altmathcal{P}}_2$. The first equality falsifies (K), since it expresses exactly the same content as the expression (\ref{1}), which is forbidden by (K).

Therefore, for \textit{irreversible} processes (K) $\Rightarrow$ (AC). A direct argument for this conclusion can be found in chapter 5 of Landsberg's book \cite{landsberg2014}. A few considerations should be made about this result. Note that the true phenomenology behind (K) in connection with (AC) is only revealed from the study of \textit{irreversible} processes. Since, if (K) dealt with the impossibility that at the end of a cycle we have $W=\altmathcal{Q} < 0$, instead of $W=\altmathcal{Q} > 0$, nothing would be changed in our analysis of the reversible case to show that (K) $\Rightarrow$ (AC). Removing this apparent mathematical ambiguity from the reversible study of (K) only occurs with the analysis of irreversible processes, revealing the true physical character of (K). To avoid overextending ourselves in this discussion, the argument concerning the reciprocal of this relationship between (AC) and (K) will not be presented here; however, it is short, and can be consulted in Dunning-Davies's paper \cite{davies1965}.

\section{Carathéodory's axiom from Clausius's principle} \label{sec:clausius}

We now present our proof of \textit{deduction of Carathéodory's axiom from Clausius's principle}. Equally as was done in the previous section, we establish the Clausius principle as Chandrasekhar \cite{chandrasekhar1939}: 

\begin{quote}[p. 24]{\cite{chandrasekhar1939}}
\textit{It is impossible that, at the end of a cycle of changes, heat has been transferred from a colder to a hotter body without at the same time converting a certain amount of work into heat.} 
\end{quote}

We shall attempt to perform a similar analysis for the Clausius principle -- which will be referred to hereafter, in abbreviated form, as (C) -- as we did for (K) in the previous section. What (C) says is that, during any thermodynamic cycle, it is not possible for a system to absorb a certain amount of heat $\altmathcal{Q}$ from a body at a lower temperature than that of the system and then fully transfer that same amount of heat $\altmathcal{Q}$ to a body at a higher temperature than that of the system, without, during this cycle, there being the conversion of some amount of work $W$ into additional heat. Here, the mentioned bodies whose system comes into contact in the described cycle are bodies that preserve their respective temperatures when interacting with the system. This makes it implicit that these bodies in contact with the system performing the cycle are thermal reservoirs. 

So, in the scheme of the Clausius principle, we have a thermal reservoir with a temperature lower than the temperature of the system, and a thermal reservoir with a temperature higher than the temperature of the system. These studied thermal reservoirs are usually given the suggestive name \textit{thermal sources}; \textit{cold source} for the thermal reservoir under lower temperature than that of the system, and \textit{hot source} for the thermal reservoir under higher temperature than that of the system. That is, suppose a system describes a thermodynamic cycle that absorbs a certain amount of heat ${\altmathcal{Q}}_{c}$ from a cold source, and then rejects another certain amount of heat ${\altmathcal{Q}}_{h}$ to a hot source, without there being in that cycle any realization of \textit{effective work} to be converted into additional heat. By (C), the following equality at the end of the cycle is impossible

\begin{equation}
|{{\altmathcal{Q}}_{c}}| = |{{\altmathcal{Q}}_{h}}|.
\label{6}
\end{equation}

That is, at the end of the cycle we cannot have equality between the magnitudes of the quantities of heat that were absorbed and rejected, respectively, from the cold source, and to the hot source. In (\ref{6}) we must write the modulus of the quantities of heat in the cycle, because, as a function of the interaction with the system, heat rejected to a source is, of course, algebraically negative. 

Thus, already familiar with the content of (AC), we wish to show that (C) $\Rightarrow$ (AC). As before, for this it is sufficient for us to show that if (AC) is false, so is (C). We will show a proof for this relation first for \textit{reversible} processes. So, let ${\altmathcal{P}}'$ be first the \textit{reversible cycle} depicted in Fig.$\,$\ref{f5} and given in the same thermodynamic space that we are already used to working in, for the same model thermodynamic system used earlier. Notice that ${\altmathcal{P}}'$ runs through the cycle of points $B \rightarrow C \rightarrow D \rightarrow A \rightarrow B$ in thermodynamic space.

\begin{figurehere}
\begin{center}	
\includegraphics[scale=0.7]{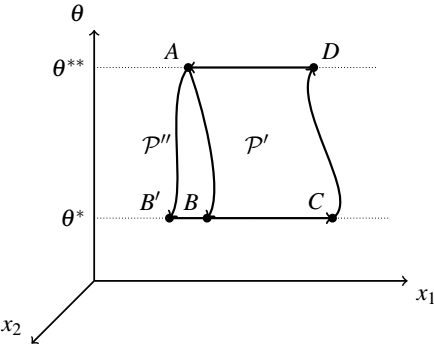}
\caption{Reversible cycle ${\altmathcal{P}}'$ constructed by intermediate processes between points $A$, $B$, $C$ and $D$ in thermodynamic space. Since ${\altmathcal{P}}'$ is reversible, the direction of travel in the cycle can be chosen arbitrarily. Here the direction $B \rightarrow C \rightarrow D \rightarrow A \rightarrow B$ is chosen. The stretch $A \rightarrow B' \rightarrow B$ and the cycle ${\altmathcal{P}}''$ will be discussed later.}
\label{f5}
\end{center}
\end{figurehere}

We construct the reversible cycle ${\altmathcal{P}}'$ from 
Fig. $\,$\ref{5}, by going through the cycle of points $B \rightarrow C \rightarrow D \rightarrow A \rightarrow B$, so that the system describes, during ${\altmathcal{P}}'$: the process $B \rightarrow C$, whose temperature ${\theta}^{*}$ remains constant during the contact of the system with the neighborhood, but there is the absorption of a certain amount of heat ${\altmathcal{Q}}_{B \rightarrow C} > 0$ of the system from the neighborhood; the process $C \rightarrow D$, which is assumed to be adiabatic, hence ${\altmathcal{Q}}_{C \rightarrow D} = 0$; the process $D \rightarrow A$, whose temperature ${\theta}^{**}$ remains constant during the contact of the system with the neighborhood, but there is the rejection of a certain amount of heat ${\altmathcal{Q}}_{B \rightarrow C} < 0$ from the system to the neighborhood; finally, the process $A \rightarrow B$, which is also supposed to be adiabatic, so ${\altmathcal{Q}}_{A \rightarrow B} = 0$. This closes the cycle ${\altmathcal{P}}'$.

Here, some important considerations should be noted: (i) in principle nothing prevents a cycle like ${\altmathcal{P}}'$ from being constructed, (ii) in general during ${\altmathcal{P}}'$ we have $|{\altmathcal{Q}}_{B \rightarrow C}| \neq |{\altmathcal{Q}}_{D \rightarrow A}|$, with \textit{effective work} being converted into additional heat, iii) the intermediate processes $B \rightarrow C$ and $D \rightarrow A$ of ${\altmathcal{P}}'$, by preserving the temperature of the system in contact with its neighborhood, indicate that during these processes the system is in contact with \textit{thermal sources}, where naturally the \textit{cold source} is that at temperature ${\theta}^{*}$ and the \textit{hot source} is that at temperature ${\theta}^{**}$, and iv) since ${\altmathcal{P}}'$ is reversible, this whole idealized construction for ${\altmathcal{P}}'$ could be reversed by reversing the cycle and taking in reverse an absorption of heat from the hot source, and a rejection of heat to the cold source.

Next, we take note of the cycle ${\altmathcal{P}}''$, which can also be seen in Fig.$\,$\ref{f5} and runs through the cycle of points $B' \rightarrow B \rightarrow C \rightarrow D \rightarrow A \rightarrow B'$ in thermodynamic space. In ${\altmathcal{P}}''$ the point $B'$ has been chosen such that we have $|{\altmathcal{Q}}_{B' \rightarrow C}| = |{\altmathcal{Q}}_{D \rightarrow A}|$ during the realization of ${\altmathcal{P}}''$, with the intermediate process $A \rightarrow B'$ also assumed to be adiabatic. 

Given the unrestricted possibility of the occurrence of ${\altmathcal{P}}'$, if it is also true that ${\altmathcal{P}}''$ is possible along the lines of what has been constructed, then it means that the adiabatic processes $A \rightarrow B'$ and $A \rightarrow B$ are simultaneously possible. However, if $A \rightarrow B'$ and $A \rightarrow B$ are simultaneously possible, then (AC) is false. In fact, we can approximate $B'$ from $B$ abitrarily. And, if we assume ${\altmathcal{P}}'$ possible, the reversible adiabatic processes $A \rightarrow B'$ and $A \rightarrow B$ become simultaneously possible. If we also arbitrarily approximate the line of points in thermodynamic space whose temperature is ${\theta}^{*}$, to the line of points whose temperature is ${\theta}^{*}$, we would have that: \textit{arbitrarily close to $A$ there are states that can be reached from $A$ by reversible adiabatic processes, thus falsifying (AC) for reversible processes}. 

However, we can see that if we assume the falsity of (AC), the execution of ${\altmathcal{P}}''$ immediately gives us the falsity of (C). Indeed, if (AC) is false then $A \rightarrow B'$ and $A \rightarrow B$ are simultaneously possible, in particular ${\altmathcal{P}}''$ is possible, and as a consequence during ${\altmathcal{P}}''$

\begin{equation}
|{\altmathcal{Q}}_{B' \rightarrow C}| = |{\altmathcal{Q}}_{D \rightarrow A}|.
\label{7}
\end{equation}

Which provides, by the expression (\ref{7}), the same as the expression (\ref{6}), which is forbidden by (C). Hence, for \textit{reversible} processes (C) $\Rightarrow$ (AC). One would naturally expect (C) $\Rightarrow$ (AC) also for the \textit{irreversible} case. In fact, an argumentation as to the validity of the relation (C) $\Rightarrow$ (AC) for \textit{irreversible} processes follows naturally from what has already been shown with the analogous analysis that (K) $\Rightarrow$ (AC) for the \textit{irreversible} case. For this argument would require the construction of two irreversible cycles similar to ${\altmathcal{P}}'$ and ${\altmathcal{P}}''$, such that in that analogous to ${\altmathcal{P}}'$ the process $A \rightarrow B$ would be assumed to be adiabatic and irreversible, and in that analogous to ${\altmathcal{P}}''$ the process $A \rightarrow B'$ would be assumed to be adiabatic and irreversible. The other processes in these cycles would be reversible. We next would take the assumption that these irreversible constructed cycles are simultaneously possible. 

So, repeating the same analysis and verifying conclusions similar to those obtained in the irreversible case of the (K) $\Rightarrow$ (AC) relation, we would show that, in order not to violate (C), (AC) is also true for \textit{irreversible} processes. To avoid repeating these same steps and the same arguments made before, which would make the discussion here unnecessarily dull, this step will not be developed in the present paper.  

\section{Road to entropy} \label{sec:entropia}

Now, armed with the validity of the Carathéodory's axiom, deduced from the principles of Clausius and Kelvin in the previous sections, and the content of the \textit{Carathéodory's theorem}, which we shall see next, we shall show how to obtain the \textit{entropy} and the mathematical content of the \textit{second law of thermodynamics} from the Carathéodory's formalism. With this goal in mind, we will first need to talk a bit about some formal aspects of Carathéodory's formalism. Seeking to combine detail, fluidity, and didactic character in the present section, we divide it into two subsections: one, \ref{subsec:teorema}, to deal with the mathematics itself and the theorem used in Carathéodory's formalism, and another, \ref{subsec:segundalei}, to deal with a possible path\footnote{There are other possible paths for this \cite{luscombe2018,buchdahl1966,landsberg2014, davies2007}.} possible for obtaining the \textit{entropy} and the mathematical content of the \textit{second law of thermodynamics} by this formalism.

\subsection{Mathematical Requirements and Carathéodory's Theorem} \label{subsec:teorema}

The first of the formal aspects of Carathéodory's formalism that we must study is the mathematical interpretation that this formalism gives to the \textit{thermodynamic coordinates}. Quantities like the \textit{empirical temperature} $\theta$, the \textit{volume} $V$, among others, which we usually call \textit{thermodynamic coordinates} when they characterize the state of a thermodynamic system, appear in Carathéodory's formalism in a formal perspective that draws a great parallel with the concept of the generalized coordinates of classical mechanics \cite{buchdahl1966}, which characterize a mechanical system.

Other quantities related to a thermodynamic system, such as \textit{heat} $\altmathcal{Q}$, \textit{work} $W$, and the \textit{energy} $E$, are identified in Carathéodory's formalism as a kind of \textit{generalized function} of the thermodynamic coordinates. Mathematically, the thermodynamic coordinates are expressed as quantities ${x}_{i}$, where the index $i$ varies according to the number of thermodynamic coordinates under analysis. For the generalized functions of the thermodynamic coordinates we have announced, on the other hand, we write that they are functions ${{\chi}_{i}}={{\chi}_{i}}({x}_{j})$, where the index $j$ tells us that the ${\chi}_{i}$ are not necessarily functions that depend on all the thermodynamic coordinates considered.

And it is precisely because of this fact that the ${\chi}_{i}$ are not always \textit{state functions} in the sense of characterizing the state of a thermodynamic system. Since they do not necessarily contain a dependence with all the thermodynamic coordinates that define that state. An example of a function ${\chi}_{i}$ that is actually a state function is the energy $E$ of a thermodynamic system, since it has dependence with all the thermodynamic coordinates that define the state of that system. An example of a ${\chi}_{i}$ function that is not a state function is the heat $\altmathcal{Q}$, which is related to the interactions of the thermodynamic system with its neighborhood and thus has no dependence with all the thermodynamic coordinates that define the state of the system. 

Already notice here the physical distinction that this formalism provides by telling us mathematically that energy is a function of state of the system and therefore is directly linked to the characterization of its state, while heat is not. In other words, the introduction of these ideas already makes it clear that a thermodynamic system can possess energy, but it cannot possess heat, since the latter depends on the thermodynamic process carried out. Moving on in this discussion, what we often see in classical thermodynamics is

\begin{equation}
{\delta}{\chi}^{*}=\sum_{i=1}^{n} {{\chi}_{i}}({x}_{j})d{{x}_{i}}.
\label{8}
\end{equation}

Careful analysis of the expression (\ref{8}) is crucial to what follows\footnote{An elegant name for the quantities expressed by (\ref{8}) is that of {\textit{pfaffian differential forms}} \cite{sneddon2006}. However, because this is merely a technical term, and would add little to the discussion of the material in this paper, we do not use this nomenclature. For the reader who wants to delve deeper into the rich mathematical theory of pfaffian differential forms, we suggest consulting Sneddon's book \cite{sneddon2006}.}. What (\ref{8}) says is that the sum of the product between the generalized functions ${\chi}_{i}$ and the infinitesimals of the thermodynamic coordinates gives us an infinitesimal of some particular generalized function ${\chi}^{*}$. That is, in classical thermodynamics the infinitesimals of ${\chi}_{i}$ are given by expressions analogous to the one in (\ref{8}). Turning to the symbols used in (\ref{8}), both ${\delta}$ and $d$ refer to infinitesimal quantities. But, as usual in physics, for an infinitesimal quantity that characterizes an \textit{exact differential}\footnote{i.e., an infinitesimal quantity whose variation only depends on its initial and final values.} we reserve the symbol $d$. Otherwise, we give the symbol ${\delta}$ for an infinitesimal quantity that is not necessarily an exact differential, and we call it an \textit{inexact differential}.

If the quantity ${\delta}{\chi}^{*}$ is an exact differential, we replace the symbol ${\delta}$ with the usual $d$ and have that the generalized function ${\chi}^{*}$ is actually a state function, having dependence with all thermodynamic coordinates ${x}_{i}$. This case makes the expression (\ref{8}) more familiar, when we have

\begin{equation}
d{\chi}^{*}=\sum_{i=1}^{n} \frac{\partial{\chi}^{*}}{\partial{x}_{i}}d{{x}_{i}}.
\label{9}
\end{equation}

That is, in this case the ${\chi}^{i}$ are the partial derivatives of ${\chi}^{*}$ with respect to the thermodynamic coordinates ${x}_{i}$ and the state function ${\chi}^{*}$ can be obtained via a ordinary integration\footnote{Com ${\delta}{\chi}^{*}$ by defining an exact differential $d{\chi}^{*}$, taking its integral need not specify any particular set of values of the thermodynamic coordinates, which would define a path for the integration to be done. In that case, the integration of $d{\chi}^{*}$ evaluates only the initial and final values of ${\chi}^{*}$.} from the expression (\ref{8}). Again, exemplifying, and from what has been previously discussed, we have as an immediate example of a thermodynamic quantity that defines an exact differential, energy, and as one that doesn't, heat. A test that we can always apply to check whether the quantity ${\delta}{\chi}^{*}$ is an exact differential or not is to evaluate, in the generalized functions ${\chi}_{i}$ of ${\delta}{\chi}^{*}$ in (\ref{8}), whether

\begin{equation}
\frac{\partial{\chi}_{k}}{\partial{x}_{l}}=\frac{\partial{\chi}_{l}}{\partial{x}_{k}} \quad \textrm{with} \quad (k,l=1,2,...,n).
\label{10}
\end{equation}

The test of equation (\ref{10}) evaluates whether the crossed partial derivatives of any pairs (${\chi}_{k},{\chi}_{l}$) of the ${\chi}_{i}$, to the corresponding pair of the thermodynamic coordinates (${x}_{l},{x}_{k}$) with naturally $(k,l=1,2,...,n)$, are identically equal to each other. The reader may recognize that this test follows from the \textit{Clairaut-Schwarz theorem}, from the \textit{differential calculus of several variables}. If equation (\ref{10}) is satisfied for any pair of indices, $(k,l=1,2,...,n)$, the quantity ${\delta}{\chi}^{*}$ will be an exact differential.

Next, we also want to pay special attention to the important situation in which the expression (\ref{8}) nullifies, that is, when

\begin{equation}
{\delta}{\chi}^{*}=\sum_{i=1}^{n} {{\chi}_{i}}({x}_{j})d{{x}_{i}}=0.
\label{11}
\end{equation}

For the equation defined in (\ref{11}) we give here the name \footnote{These equations are customarily called {\textit{Pfaff differential equations}} \cite{sneddon2006}.} of \textit{differential equation associated with ${\delta}{\chi}^{*}$}. It is important for us to note that the solutions of these differential equations associated with ${\delta}{\chi}^{*}$ will always be a set of thermodynamic coordinate values, which can be visualized, in the geometric perspective of thermodynamic space, as a set of points given by these thermodynamic coordinate values in that space. In particular, when ${\delta}{\chi}^{*}$ is an exact differential, that set of points in thermodynamic space forming the solutions of the differential equation associated with ${\delta}{\chi}^{*}=0$ is, from equation (\ref{11})

\begin{equation}
{\chi}^{*}={\chi}^{*}({x}_{1},{x}_{2},...,{x}_{n})=c.
\label{12}
\end{equation}

Where in equation (\ref{12}) $c$ is a constant. Equation (\ref{12}) is a \textit{hypersurface} of \textit{n dimensions} in thermodynamic space with $n$ thermodynamic coordinates. Then, fixed a hypersurface ${\chi}^{*}({x}_{1},{x}_{2},...,{x}_{n})=c$, for a given value of $c$, one naturally establishes the solutions of (\ref{12}), which are \textit{hypercurves} in this space. At first abstract, this conclusion materializes much more intuitively when we work with three thermodynamic coordinates and then the equation (\ref{12}) translates into a familiar \textit{surface} in \textit{3 dimensions}, ${\chi}^{*}({x}_{1},{x}_{2},{x}_{3})=c$, in the related three-dimensional thermodynamic space. Furthermore, the solutions of ${\chi}^{*}({x}_{1},{x}_{2},{x}_{3})=c$ become \textit{curves} in this space. It is worth noting that it is from three thermodynamic coordinates that we usually model and study most thermodynamic systems from classical thermodynamics at the undergraduate level.

Finally, before we talk about Carathéodory's theorem, we need to deal with when the quantity ${\delta}{\chi}^{*}$ in (\ref{8}) does not constitute an exact differential, but becomes one when it is multiplied by a certain function, which we call the \textit{integrating factor}. In fact, analogous to ${\chi}_{i}$, let ${\eta}$ be any generalized function of thermodynamic coordinates which at first is not an exact differential, so the infinitesimal of ${\eta}$ is ${\delta}\eta$. It turns out that in some cases, when we multiply ${\delta}\eta$ by some other generalized thermodynamic coordinate function ${\mu}$, we get the validity of the following relation

\begin{equation}
d\sigma={\mu}{\delta}\eta.
\label{13}
\end{equation}

That is, in some cases, we can multiply an inexact differential ${\delta}\eta$ by an appropriate generalized function of the thermodynamic coordinates $\mu$, so that we get an exact differential $d\sigma$. When this happens, we say that ${\delta}\eta$ is an \textit{integrable inexact differential} by $\mu$, hence the suggestive name \textit{integrating factor} for $\mu$. This action, given in (\ref{13}), is also called \textit{integrate the equation} differential associated with $\delta\eta$. And $\delta\eta$ being a \textit{integrable inexact differential}, the solutions of its respective associated differential equation are also of the form of the expression (\ref{12}). But in which cases can we do this \textit{integration}? Except for the cases restricted to two or three thermodynamic coordinates in study\footnote{Suggest reading Buchdahl's book \cite{buchdahl1966} for the reader who wishes to study these simpler cases in depth.}, the result that generalizes the answer to this question, and gives physical meaning to all this previous mathematical construction, is the \textit{Carathéodory's theorem}.

It then follows Carathéodory's theorem, adapted to the notation of the present paper, from what appears in H. A. Buchdahl's \cite{buchdahl1966} book of classical thermodynamics \textit{The Concepts of Classical Thermodynamics}:

\begin{quote}[p. 62]{\cite{buchdahl1966}}
\textit{If every neighbourhood of any arbitrary point A contains points B inaccessible from A along solutions curves of the equation $\sum_{i=1}^{n} {{\chi}_{i}}({x}_{j})d{{x}_{i}}=0$, then the equation is integrable.}
\end{quote}

Before we prove Carathéodory's theorem, let's make evident to the reader the familiar physical content it has from Carathéodory's axiom, which we have already discussed extensively in the \ref{sec:kelvin} and \ref{sec:clausius} sections. First, let's recapitulate that the heat $\altmathcal{Q}$ related to a thermodynamic system is here mathematically identified as a generalized function of thermodynamic coordinates according to (\ref{8}), so for $n$ any ${x}_{i}$ thermodynamic coordinates, we have

\begin{equation}
{\delta}{\altmathcal{Q}}=\sum_{i=1}^{n} {{\altmathcal{Q}}_{i}}({x}_{j})d{{x}_{i}}.
\label{14}
\end{equation}

Naturally, with the corresponding \textit{differential equation associated with $\altmathcal{Q}$}, given by

\begin{equation}
{\delta}{\altmathcal{Q}}=\sum_{i=1}^{n} {{\altmathcal{Q}}_{i}}({x}_{j})d{{x}_{i}}=0.
\label{15}
\end{equation}

Note that equation (\ref{15}) already gives us, precisely, the mathematical description of a \textit{infinitesimal adiabatic process}. Now, if we translate and apply Carathéodory's theorem to $\altmathcal{Q}$, and to points in the thermodynamic space of any $n$ thermodynamic coordinates we want to study, we can write that: \textit{if any neighborhood of any point A contains points B that are inaccessible by A from the solution curves of the equation ${\delta}{\altmathcal{Q}}=0$, then this equation is integrable.}

But, we note here a congruence between the premise of Carathéodory's theorem applied to $\altmathcal{Q}$, and Carathéodory's axiom, which we remind the reader again, already translating it to the context of thermodynamic space\footnote{Remember, we have already justified the univocal relation between thermodynamic states and points in thermodynamic space.}: 

\noindent
\textbf{Carathéodory's axiom.} \textit{arbitrarily close to any given point there are points that are inaccessible from an initial point by means of adiabatic processes.}\\

And, as we also recall, adiabatic processes are processes in which the equation (\ref{15}) occurs throughout its execution. Also, realize that an adiabatic process is exactly the physical concept that is mathematically represented in this formalism by means of the solution curves of the equation (\ref{15}). In other words, \textit{adiabatic processes are the solution curves of the equation} (\ref{15}). Note the elegant distinction, and at the same time connection, that Carathéodory's formalism makes between the physical substance and the mathematical substance of classical thermodynamics.

Thus, we rewrite Carathéodory's theorem as follows: \textit{if arbitrarily close to any given point there are points inaccessible from an initial point by adiabatic processes, then the equation ${\delta}{\altmathcal{Q}}=0$ is integrable.} Or, to summarize:\\

\noindent
\textbf{Carathéodory's theorem.} \textit{if Carathéodory's axiom holds, then ${\delta}{\altmathcal{Q}}=0$ is integrable.}\\

Carathéodory's theorem speaks in mathematical terms when an equation of type (\ref{11}) is integrable, and Carathéodory's axiom points out in physical terms that heat always is. For mere initial simplicity, we prove Carathéodory's theorem for any three thermodynamic coordinates $({{x}_{1}},{{x}_{2}},{{x}_{3}})$. As we had announced earlier, we follow the arguments of Born \cite{born1921}.\\

\noindent
\textit{Proof.} Consider that, as our initial hypothesis, the Carathéodory's axiom holds. That is, consider that an arbitrary point B in 3-dimensional thermodynamic space is inaccessible from an also arbitrary point A, however close B and A are, by the solutions of ${\delta}{\altmathcal{Q}}=0$ in that space. Let us now say that there is a second arbitrary point C accessible to A by ${\delta}{\altmathcal{Q}}=0$. Then A and C are accessible to each other by the solutions of ${\delta}{\altmathcal{Q}}=0$. But C must also be inaccessible to B by these solutions of ${\delta}{\altmathcal{Q}}=0$, because otherwise, through passing through C, B would be accessible to A, which contradicts our initial hypothesis. Therefore, the points accessible to A define a surface in the thermodynamic space containing A such that this surface also contains all the points accessible to A by the solutions of ${\delta}{\altmathcal{Q}}=0$. Now, since this property of inaccessibility of points in space was exemplified by A, with arbitrary A, it must also hold for any other point in that thermodynamic space. Thus defining, for each point chosen, and by the fact that there are always points arbitrarily close to the point one chooses, a surface in three-dimensional space, $\sigma({{x}_{1}},{{x}_{2}},{{x}_{3}})=c$, which contains all the respective points accessible from the arbitrary point chosen. There are thus, as a consequence of our inaccessibility hypothesis, and of there being inaccessible points arbitrarily close to the point that is chosen, several neighboring surfaces that do not intersect\footnote{This verification is simple, as Landsberg discusses \cite{landsberg2014}.}, $\Sigma({{x}_{1}},{{x}_{2}},{{x}_{3}})=c$, containing the points that are accessible to each other. On these surfaces we must have $d\sigma=0$ and ${\delta}{\altmathcal{Q}}=0$, from which we conclude that ${\delta}{\altmathcal{Q}}$ and $d\sigma$ must be proportional quantities on these surfaces. That is, there exists ${\mu}={\mu}({{x}_{1}},{{x}_{2}},{{x}_{3}})$ such that

\begin{equation}
d\sigma={\mu}{\delta}{\altmathcal{Q}}.
\label{16}
\end{equation}
\QEDA

For the general case of $n$ thermodynamic coordinates the above argument is the same, exchanging only the term and the construction of \textit{surfaces} for the generalization of \textit{hypersurfaces}. It follows from the above proof that writing ${\mu}$ as a quantity that relates to the differentials $d\sigma$ and ${\delta}{\altmathcal{Q}}$ exactly as in the form expressed in equation (\ref{16}) is not mandatory. In other words, according to Carathéodory's theorem, mathematically we only need ${\mu}$ to express the proportionality that exists between $d\sigma$ and ${\delta}{\altmathcal{Q}}$. Thus, for reasons that will become clear later, we will write for ${\delta}{\altmathcal{Q}}$ and ${d\sigma}$, instead of the equation (\ref{16}), the following expression 

\begin{equation}
{\delta}{\altmathcal{Q}}={\mu}d\sigma.
\label{17}
\end{equation}

Note at this point one of the striking features of Carathéodory's formalism: ambiguities and general mathematical conclusions will necessarily be subject to physical evaluations of their meaning. Let us now, in order to better understand the equality (\ref{17}) and obtain the \textit{entropy} and the mathematical content of the \textit{second law of thermodynamics} by the formalism developed here, seek to better study the integral factor ${\mu}$ and the state function $\sigma$ that we have just discovered.

\subsection{Entropy and absolute temperature} \label{subsec:segundalei}

In order to obtain the \textit{entropy} and the mathematical content of the \textit{second law of thermodynamics} by the Carathéodory formalism, we must analyze the situation of two thermodynamic systems in purely thermal contact with each other, as well as adiabatically isolated from the surrounding neighborhood. And at this point, we shall again deal with any $n$ number of thermodynamic coordinates for the thermodynamic systems under study, assuming the validity of Carathéodory's theorem for this general case as well. Consider then two thermodynamic systems, which we will refer to as $K_{A}$ and $K_{B}$, in purely thermal contact with each other, i.e., $K_{A}$ and $K_{B}$ can interact only via heat exchange. Furthermore, as we said, $K_{A}$ and $K_{B}$ are adiabatically isolated from their surrounding neighborhood. 

If we look at the set $K_{A}$ and $K_{B}$ globally, we can say that both systems form a compound thermodynamic system $K_{C}$. Looking at this composite system $K_{C}$, in mathematical terms, we have

\begin{equation}
{\delta}{{\altmathcal{Q}}_{C}}={\delta}{{\altmathcal{Q}}_{A}}+{\delta}{{\altmathcal{Q}}_{B}}.
\label{18}
\end{equation}

When thermal equilibrium is established between $K_{A}$ and $K_{B}$, both individual systems will have the same empirical temperature $\theta$ at the end of a given time\footnote{In general, the time for thermodynamic equilibrium to occur during any thermodynamic interaction -- mechanical, chemical, etc. -- between thermodynamic systems is called the relaxation time. A beautiful discussion of this concept in its macroscopic aspect can be found in Callen's book \cite{callen1985}.}, which will also naturally be the same as the empirical temperature of the composite system $K_{C}$ at equilibrium. Now, being $({x}_{1},{x}_{2},...,{x}_{n-1},{\theta}_{A})$, $({y}_{1},{y}_{2},...,{y}_{n-1},{\theta}_{B})$, e $({x}_{1},{x}_{2},...,{x}_{n-1},{y}_{1},{y}_{2},... ,{y}_{n-1},{\theta}_{A},{\theta}_{B})$, the respective $n$ thermodynamic coordinates of $K_{A}$, $K_{B}$ and $K_{C}$, we will have, at thermal equilibrium, ${\theta}_{A}={\theta}_{B}={\theta}$. The coordinates $x_i$ and $y_i$ are the thermodynamic coordinates that provide the mechanical behavior of the respective systems $K_{A}$ and $K_{B}$. These findings will be important later on. By applying equation (\ref{17}) to equation (\ref{18}), we immediately have

\begin{equation}
{{\mu}_{C}}{d\sigma_{C}}={{\mu}_{A}}{d\sigma_{A}}+{{\mu}_{B}}{d\sigma_{B}}.
\label{19}
\end{equation}

\noindent
Rearranging (\ref{19}) in terms of $d\sigma_{C}$, we obtain

\begin{equation}
{d\sigma_{C}}=\frac{{\mu}_{A}}{{\mu}_{C}}{d\sigma_{A}}+\frac{{\mu}_{B}}{{\mu}_{C}}{d\sigma_{B}}.
\label{20}
\end{equation}

As justified by Carathéodory's theorem, the ${\sigma}$ quantities that arise from the integration of ${\delta}{\altmathcal{Q}}$ quantities are state functions, and thus are functions of all thermodynamic coordinates that characterize the state of a thermodynamic system. From the expression (\ref{20}), we also see that ${\sigma}_{C}={\sigma}_{C}({\sigma}_{A},{\sigma}_{B})$. That is, we can write

\begin{equation}
{d\sigma_{C}}=\frac{\partial {{\sigma}_{C}}}{\partial {{\sigma}_{A}}}{d\sigma_{A}}+\frac{\partial {{\sigma}_{C}}}{\partial {{\sigma}_{B}}}{d\sigma_{B}}.
\label{21}
\end{equation}

\noindent
So, comparing the expressions (\ref{21}) and (\ref{20}), we have

\begin{subequations} \label{22}
\begin{align}
\frac{{\mu}_{A}}{{\mu}_{C}}=\frac{\partial {{\sigma}_{C}}}{\partial {{\sigma}_{A}}}; \label{22a}\\
\frac{{\mu}_{B}}{{\mu}_{C}}=\frac{\partial {{\sigma}_{C}}}{\partial {{\sigma}_{B}}}. \label{22b}
\end{align}
\end{subequations}

\noindent
The equations (\ref{22}) tell us that the quotients $\frac{{\mu}_{A}}{{\mu}_{C}}$ and $\frac{{\mu}_{B}}{{\mu}_{C}}$ are such that

\begin{subequations} \label{23}
\begin{align}
\frac{{\mu}_{A}}{{\mu}_{C}}=\frac{{\mu}_{A}}{{\mu}_{C}}({{\sigma}_{A}},{{\sigma}_{B}}); \label{23a}\\
\frac{{\mu}_{B}}{{\mu}_{C}}=\frac{{\mu}_{B}}{{\mu}_{C}}({{\sigma}_{A}},{{\sigma}_{B}}). \label{23b}
\end{align}
\end{subequations}

But, we know that in principle, by Carathéodory's theorem, the quantities $\mu$ are functions of the thermodynamic coordinates defining the states of their respective parent thermodynamic systems, i.e., and already for the thermal equilibrium between $K_{A}$ and $K_{B}$

\begin{subequations} \label{24}
\begin{align}
{\mu}_{A}={\mu}_{A}({x}_{1},{x}_{2},...,{x}_{n-1},{\sigma}_{A},{\theta}); \label{24a}\\
{\mu}_{B}={\mu}_{B}({y}_{1},{y}_{2},...,{y}_{n-1},{\sigma}_{B},{\theta}); \label{24b}\\
{\mu}_{C}={\mu}_{C}({x}_{1},...,{x}_{n-1},{y}_{1},...,{y}_{n-1},{\sigma}_{A},{\sigma}_{B},\theta). \label{24c}
\end{align}
\end{subequations}

Thus, pay attention that to reconcile equations (\ref{24}) and (\ref{23}) the quantities $\mu$ should not depend on their respective thermodynamic coordinates that provide the mechanical behavior of the respective system related to $\mu$. Otherwise, the quotients given in the equations (\ref{23}) could not depend on the $\sigma$ quantities alone. Hence, by this analysis, the most general form of the $\mu$ quantities must be\footnote{A more detailed argument for this step can be found in Zemansky's paper \cite{zemansky1966}.}

\begin{subequations} \label{25}
\begin{align}
{\mu}_{A}={\mu}_{A}({\sigma}_{A},{\theta})=t(\theta)f_{A}({\sigma}_{A}); \label{25a}\\
{\mu}_{B}={\mu}_{B}({\sigma}_{B},{\theta})=t(\theta)f_{B}({\sigma}_{B}); \label{25b}\\
{\mu}_{C}={\mu}_{C}({\sigma}_{A},{\sigma}_{B},{\theta})=t(\theta)f_{C}({\sigma}_{A},{\sigma}_{B}). \label{25c}
\end{align}
\end{subequations}

Note that, only thanks to the thermal equilibrium between the systems $K_{A}$ and $K_{B}$ are we able to write the equations (\ref{25}), which show that the integral factors $\mu$ can be written with respect to a function of universal character\footnote{We should point out that the Clausius formalism also indicates that temperature is related to the integral factor of heat \cite{zemansky1997}. In the Gibbs formalism, although also present, this fact is not very relevant.} $t=t(\theta)$, independent of the thermodynamic system, which depends solely on the empirical temperature $\theta$ of the equilibrium between $K_{A}$ and $K_{B}$ and which is therefore common to both systems $K_{A}$ and $K_{B}$. 

The universal content of this function $t=t(\theta)$ motivates us to define the so-called \textit{ absolute temperature} $T$ of \cite{pauli1973} thermodynamic systems, minus an arbitrary constant $k$, so that

\begin{equation}
T=kt(\theta).
\label{26}
\end{equation}

The algebraic sign of the arbitrary constant $k$ defining the absolute temperature in equation (\ref{26}) \textit{is a convention}\footnote{We strongly recommend the reader to consult section 11 and appendices [A-6] and [A-8] of Pauli's book \cite{pauli1973}, for an enlightening discussion of this subject.} and Carathéodory's formalism makes this clear. The historical choice was $k>0$. Moreover, it was to obtain this direct proportionality between the temperature scales $T$ and $t(\theta)$ given in (\ref{26}) that we chose to work with equation (\ref{17}), instead of equation (\ref{16}). This suggests for the absolute temperature the same experimental methods as are used for the empirical temperature determination, as discussed by Buchdahl \cite{buchdahl1966}.

If we substitute the result of equations (\ref{25}), together with equation (\ref{26}), into equation (\ref{17}), we will have, for any thermodynamic system

\begin{equation}
{\delta}{\altmathcal{Q}}=\frac{T}{k}f(\sigma)d\sigma.
\label{27}
\end{equation}

\noindent
Isolating the terms of (\ref{27}) with the dependency on $\sigma$

\begin{equation}
\frac{\delta{\altmathcal{Q}}}{T}=\frac{1}{k}f(\sigma)d\sigma.
\label{28}
\end{equation}

We already know that the quantity $\sigma$ is a state function, as is, for example, the energy $E$ of a thermodynamic system. We then call the $\sigma$ function \textit{empirical entropy}, in allusion to its relation to the empirical temperature of a thermodynamic system. Next, we define

\begin{equation}
d\altmathcal{S}\equiv\frac{1}{k}f(\sigma)d\sigma.
\label{29}
\end{equation}

The quantity $\altmathcal{S}$ is clearly also a state function, and is called the \textit{absolute entropy}, or just the \textit{entropy} of the thermodynamic system, also alluding to its relationship with the absolute temperature. The measurement of entropy is also independent of the particular properties of each thermodynamic system. Finally, substituting equation (\ref{29}) into equation (\ref{28}), we obtain

\begin{equation}
d\altmathcal{S}=\frac{\delta{\altmathcal{Q}}}{T}.
\label{30}
\end{equation}

The expression (\ref{30}) is the mathematical content of the \textit{second law of thermodynamics} for \textit{reversible processes}. Why (\ref{30}) is only valid for reversible processes becomes clear as we try to evaluate the entropy $\altmathcal{S}$ from (\ref{30}) for irreversible processes. If we try to do this, we will immediately see that a simple integration of $d\altmathcal{S}$ is not possible to obtain $d\altmathcal{S}$ for irreversible processes. For such an integration, the absolute temperature $T$ in (\ref{30}) would not even be defined in the intermediate steps of any irreversible process we would consider evaluating. This means that we need another strategy to study $\altmathcal{S}$ in irreversible processes. 

However, fortunately, Carathéodory's axiom provides us with that other strategy quickly. To do this, consider, again for pure simplicity, a thermodynamic space with three thermodynamic coordinates, ${x}_{1},{x}_{2}$ and $\altmathcal{S}$. Then consider the schematic of this thermodynamic space in Fig.$,$\ref{f6} with two supposedly adiabatic irreversible processes, $\altmathcal{P}_{+}$ and $\altmathcal{P}_{-}$, both starting from the same arbitrary point $A$.

\begin{figurehere}
\begin{center}	
\includegraphics[scale=0.7]{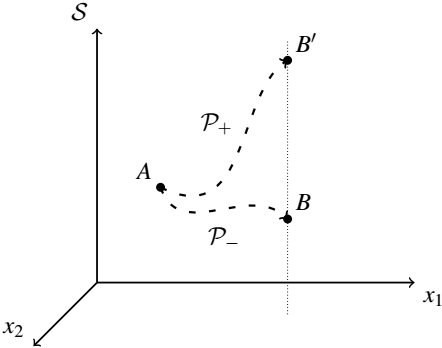}
\caption{Thermodynamic space of the coordinates ${x}_{1},{x}_{2}$ and $\altmathcal{S}$, with the assumed irreversible adiabatic processes $\altmathcal{P}_{+}$ of $A \rightarrow B'$, and $\altmathcal{P}_{-}$ of $A \rightarrow B$.}
\label{f6}
\end{center}
\end{figurehere}

That done, suppose both of these processes, $\altmathcal{P}_{+}$ and $\altmathcal{P}_{-}$, simultaneously possible. Then, the entropy variation in the course of these two processes could be, from Fig.$\,$\ref{f6}, either positive or negative. Positive during $\altmathcal{P}_{+}$, negative during $\altmathcal{P}_{-}$. But, we can arbitrarily approximate $B'$ from $B$, and since we assume $\altmathcal{P}_{+}$ and $\altmathcal{P}_{-}$ simultaneously possible, it follows that all points on the line containing $B$ and $B'$ can be reached from $A$ by irreversible adiabatic processes.  Also approximating the line containing $B$ and $B'$ from $A$, we will have that: \textit{arbitrarily close to $A$ there are states that can be reached from $A$ by irreversible adiabatic processes, thus falsifying Carathéodory's axiom for irreversible processes}.

The following is derived from this argument: \textit{to not violate Carathéodory's axiom, the \textit{entropy} of any thermodynamic system in the course of an irreversible adiabatic process must always either increase or decrease, never both} \cite{pauli1973}. Again here, and perhaps even more importantly, Carathéodory's formalism impels us to note the distinction between the physical substance of classical thermodynamics and the mathematical construct developed to describe it. That is, the choice for the algebraic sign of the entropy variation in irreversible adiabatic processes becomes \textit{arbitrary}. As the reader may already know, we choose the positive sign for this variation and write

\begin{equation}
\Delta{{\altmathcal{S}}_{\altmathcal{Q}=0}}>0.
\label{31}
\end{equation}

The expression in (\ref{30}) is the mathematical content of the \textit{second law of thermodynamics} for \textit{irreversible processes}. Putting the expressions (\ref{30}) and (\ref{31}) together, when we take an adiabatic process in (\ref{30}), we get

\begin{equation}
\Delta{{\altmathcal{S}}_{\altmathcal{Q}=0}}\geqslant0.
\label{32}
\end{equation}

The expression (\ref{32}) is the general mathematical content of the \textit{second law of thermodynamics}. 

Finally, we would like the reader to take the time to observe the essence of Carathéodory's formalism. In it, despite the mathematical investment that is required, the physics of classical thermodynamics is largely reflected and requested, while the mathematical construction accurately distinguishes the physical content of the theory from the purely formal and abstract. Carathéodory's formalism still provides a vast scope for pedagogical connection between classical thermodynamics and classical mechanics, starting with the concept of generalized coordinates, as Buchdahl \cite{buchdahl1966} points out. Also, by being able to reproduce the results of classical thermodynamics simply by assuming the validity of the Carathéodory axiom at first, we draw a parallel between this axiom and the one found in Hamilton's formalism -- \textit{Hamilton's principle} --, again in the context of classical mechanics, as Pippard says \cite{pippard1966}. In other words, despite the misfortune of the unpopularity of the Carathéodory formalism over the years, this approach can definitely add much to the teaching of classical thermodynamics, and can be used as a viable alternative to the Clausius and Gibbs formalisms.

\section{Conclusion} \label{sec:conclusoes}

In this paper, we cover some of the construction of the Carathéodory formalism for classical thermodynamics in relation to the other best known formalisms of that theory. In the section \ref{sec:formalismo} we discuss some of the origin and tenor of the Carathéodory formalism in the context of its historical unpopularity over the years. We advocate for Carathéodory's formalism by fostering it as a viable alternative to teaching classical thermodynamics.

In furtherance of this cause, we seek to didactically show the reader how Carathéodory's axiom of the second law of thermodynamics can be deduced from the principles of Clausius and Kelvin: we show (K) $\Rightarrow$ (AC) in the \ref{sec:kelvin} section, and (C) $\Rightarrow$ (AC) in the \ref{sec:clausius} section. From the extensive literature reviewed, we have given a new proof for (C) $\Rightarrow$ (AC). 

In addition, we guide the reader in the \ref{sec:entropia} section through one of the possible paths that lead to obtaining the \textit{entropy} and the mathematical content of the \textit{second law of thermodynamics} from this formalism. Thus, we hope that this work will serve in some measure to popularize the Carathéodory formalism in disciplines of classical thermodynamics at the undergraduate level, also contributing to the teaching of classical thermodynamics itself. 

\section*{Acknowledgments}

Thanks are due to Paulo Henrique Ribeiro Peixoto and Gustavo Camelo Neto, professors at UFPE.

\end{multicols}

\end{document}